\documentclass[aoas,nameyear,seceqn,rotating,dvips]{arximspdf}
\usepackage{dcolumn}
\usepackage{graphicx}

\doi{10.1214/10-AOAS345}
\volume{4}
\issue{4}
\pubyear{2010}
\firstpage{1698}
\lastpage{1721}

\makeatletter

\newtheorem{proposition}{Proposition}
\newcolumntype{d}[1]{D{.}{.}{#1}}
\newcolumntype{a}[1]{D{,}{,}{#1}}
\newcolumntype{b}[1]{D{,}{\ }{#1}}
\newcolumntype{e}[1]{D{\ }{}{#1}}
\newcolumntype{f}[1]{D{,}{\ }{#1}}
\newproclaim{example}{Example}
\makeatother

\begin{document}

\begin{frontmatter}

\title{Multicategory vertex discriminant analysis for high-dimensional data}
\runtitle{Multicategory VDA for High-Dimensional Data}

\begin{aug}
\author[A]{\fnms{Tong Tong} \snm{Wu}\thanksref{Tx1}\corref{}\ead[label=e1]{ttwu@umd.edu}}
\thankstext{Tx1}{Supported
by NSF Grant CCF-0926194 and General Research Board (GRB) Award and Research
Support Award from the University of Maryland, College Park.}
\and
\author[B]{\fnms{Kenneth} \snm{Lange}\thanksref{Tx2}\ead[label=e2]{klange@ucla.edu}}
\thankstext{Tx2}{Supported in part by NIH Grants GM53275 and MH59490.}
\runauthor{T. T. Wu and K. Lange}
\affiliation{University of Maryland and University of California}
\address[A]{Department of Epidemiology \\
\quad and Biostatistics \\
University of Maryland \\
1242C SPH Building \\
College Park, Maryland 20707\\
USA\\ \printead{e1}} 

\address[B]{Departments of Biomathematics \\
\quad and Human Genetics \\
Geffen School of Medicine at UCLA \\
Gonda Research Center \\
695 Charles E. Young Drive South \\
Los Angeles, California 90095-7088\\
USA\\ \printead{e2}}

\end{aug}

\received{\smonth{7} \syear{2009}}
\revised{\smonth{3} \syear{2010}}

\begin{abstract}
In response to the challenges of data mining, discriminant analysis
continues to evolve as a vital branch of statistics.
Our recently introduced method of vertex discriminant analysis (VDA) is
ideally suited to handle multiple categories and
an excess of predictors over training cases. The current paper explores
an elaboration of VDA that conducts classification
and variable selection simultaneously. Adding lasso ($\ell_1$-norm) and
Euclidean penalties to the VDA loss function eliminates
unnecessary predictors. Lasso penalties apply to each predictor
coefficient separately; Euclidean penalties group the collective
coefficients of a single predictor. With these penalties in place,
cyclic coordinate descent accelerates estimation of all coefficients.
Our tests on simulated and benchmark real data demonstrate the virtues
of penalized VDA in model building and prediction in high-dimensional settings.
\end{abstract}


\end{frontmatter}

\section{Introduction}

Despite its long history, discriminant analysis is still undergoing
active development.
Four forces have pushed classical methods to the limits of their
applicability: (a) the sheer scale of modern datasets,
(b) the prevalence of multicategory problems, (c) the excess of
predictors over cases, and (d)
the exceptional speed and memory capacity of modern computers.
Computer innovations both solve and drive
the agenda of data mining. What was unthinkable before has suddenly
become the focus of considerable mental energy.
The theory of support vector machines (SVM) is largely a response to
the challenges of binary classification.
SVMs implement a geometric strategy of separating two classes by an
optimal hyperplane.
This simple paradigm breaks down in passing from two classes to
multiple classes.
The one-versus-rest (OVR) remedy reduces classification with $k$
categories to binary classification.
Unfortunately, OVR can perform poorly when no dominating class exists [\citet{lee04}].
The alternative of performing all $k \choose 2$ pairwise comparisons
[\citet{kressel99}]
has value, but it constitutes an even more egregious violation of the
criterion of parsimony.
In the opinion of many statisticians, simultaneous classification is
more satisfying theoretically and practically.
This attitude has prompted the application of hinge loss functions in
multicategory SVM [\citet
{bredensteiner99}; \citet{crammer01}; \citet{guermeur02}; \citet{lee04}; Liu, Shen and Doss (\citeyear{liu05}, \citeyear{liu06});
\citet{liu07}; \citet{vapnik98}; \citet{weston99}; \citet{zhang04b}; \citet{zou06}; \citet{yuan09}].

Our earlier paper [\citet{lange08}] introduced a new method of
multicategory discriminant analysis that shares many of the attractive
properties of multicategory SVM under hinge loss. These properties
include simple linear prediction of class vertices, creation of dead
regions where predictions incur no loss, and robustness to outliers.
Our vertex discriminant analysis (VDA) procedure has the advantage of
operating in ($k-1$)-dimensional space rather than in $k$-dimensional
space. Each class is represented by a vertex of a regular simplex, with
the vertices symmetrically arranged on the surface of the unit ball in
$\mathbb{R}^{k-1}$. These conventions emphasize symmetry, eliminate
excess parameters and constraints, and simplify computation and model
interpretation. For hinge loss we substitute $\varepsilon$-insensitive
loss. Both loss functions penalize errant predictions; the difference
is that hinge loss imposes no penalty when wild predictions fall on the
correct side of their class indicators. A generous value of $\varepsilon$
makes $\varepsilon$-insensitive loss look very much like hinge loss. In
addition, $\varepsilon$-insensitive loss enjoys a computational advantage
over hinge loss in avoiding constraints. This makes it possible to
implement rapid coordinate descent. Class assignment in VDA is defined
by a sequence of conical regions anchored at the origin and surrounding
the class vertices.

Modern methods of discriminant analysis such as VDA are oriented to
data sets where the number of predictors $p$ is comparable to or larger
than the number of cases $n$. In such settings it is prudent to add
penalties that shrink parameter estimates to 0. Our paper
[\citet{lange08}]
imposes a ridge penalty to avoid overfitting. Although
shrinkage forces a predicted point toward the origin, the point tends
to stay within its original conical region. Hence, no correction for
parameter shrinkage is needed, and the perils of underprediction are
mitigated. A ridge penalty also adapts well to an MM algorithm for
optimizing the loss function plus penalty.

Motivated by problems such as cancer subtype classification, where the
number of predictors $p$ far exceeds the number of observations $n$, we
resume our study of VDA in the current paper. In this setting
conventional methods of discriminant analysis prescreen predictors
[\citet{dudoit02}; \citet{li04}; \citet{li05};] before committing to a full analysis.
\citet{wang07} argue against this arbitrary step of univariate feature
selection and advocate imposing a lasso penalty. Ridge penalties are
incapable of feature selection, but lasso penalties encourage sparse
solutions. Consumers of statistics are naturally delighted to see
classification reduced to a handful of predictors. In our experience,
it is worth adding further penalties to the loss function. \citet{zhang08}
suggest $\ell_\infty$ penalties that tie together the
regression coefficients pertinent to a single predictor. In our setting
Euclidean penalties achieve the same goal and preserve spherical
symmetry. By design $\ell_1$ penalties and Euclidean penalties play
different roles in variable selection. One enforces sparsity of
individual variables, while the other enforces sparsity of grouped
variables. In the sequel we denote our original VDA with a ridge
penalty as VDA$_{\mathrm{R}}$, the modified VDA with a lasso penalty as
VDA$_{\mathrm{L}}$, the modified VDA with a Euclidean penalty as VDA$_{\mathrm{E}}$,
 and the modified VDA with both lasso and Euclidean penalties as
VDA$_{\mathrm{LE}}$. The same subscripts will be attached to the
corresponding penalty tuning constants.

A second objection to VDA$_{\mathrm{R}}$ as it currently stands is that the
computational complexity of the underlying MM algorithm scales as
$O(p^3)$. This computational hurdle renders high-dimensional problems
intractable. Although substitution of lasso penalties for ridge
penalties tends to complicate optimization of the objective function,
prior experience with lasso penalized regression
[\citet{friedman07}; \citet{wu08}]
suggests updating one parameter at a time. We
implement cyclic coordinate descent by repeated application of Newton's
method in one dimension. In updating a single parameter by Newton's
method, one can confine attention to the intervals to the left or right
of the origin and ignore the kink in the lasso. The kinks in $\varepsilon
$-insensitive loss are another matter. We overcome this annoyance by
slightly smoothing the loss function. This maneuver preserves the
relevant properties of $\varepsilon$-insensitive loss and leads to fast
reliable optimization that can handle thousands of predictors with
ease. Once the strength of the lasso penalty is determined by
cross-validation, model selection is complete.

In practice, cross-validation often yields too many false positive
predictors. This tendency has prompted \citet{meinshausen09} to
introduce the method of stability selection, which requires selected
predictors to be consistently selected across random subsets of the
data. Here we demonstrate the value of stability selection in
discriminant analysis. Because our revised versions of VDA are fast,
the 100-fold increase in computing demanded by stability selection is
manageable.

Before summarizing the remaining sections of this paper, let us mention
its major innovations: (a) the new version of VDA conducts
classification and variable selection simultaneously, while the
original VDA simply ignores variable selection; (b) coordinate descent
is substituted for the much slower MM algorithm, (c)~$\varepsilon$-insensitive
 loss is approximated by a smooth loss to accommodate
Newton's method in coordinate descent, (d) Fisher consistency is
established, (e) a grouped penalty is added, and (f) the new VDA is
tested on a fairly broad range of problems. These changes enhance the
conceptual coherence, speed, and reliability of VDA.

The rest of the paper is organized as follows. Section \ref{sec2} precisely
formulates the VDA model, reviews our previous work on VDA$_{\mathrm{R}}$,
and derives cyclic coordinate descent updates for VDA$_{\mathrm{L}}$.
Section \ref{sec3} introduces the Euclidean penalty for grouped predictors and
sketches the necessary modifications of cyclic coordinate descent.
Section \ref{Fisher_consistency_section} takes a theoretical detour and shows that $\varepsilon
$-insensitive loss is Fisher consistent. By definition, Fisher
consistent classifiers satisfy the Bayes optimal decision rule. There
is already a considerable literature extending previous proofs of
Fisher consistency from binary classification to multicategory
classification [\citet{lee04}; \citet{liu06}; \citet{liu07}; \citet{wang07}; \citet{zhang04a}; \citet{zou08}]. Section
\ref{sec5} quickly reviews the basics of stability selection. Sections \ref{sec6} and
\ref{sec7}
report our numerical tests of VDA on simulated and real data. Section
\ref{sec8}
concludes with a brief summary and suggestions for further research.

\section{Modified vertex discriminant analysis}\label{sec2}

\subsection{Ridge penalized vertex discriminant analysis (VDA$_{\mathrm R}$)}

Vertex discriminant analysis (VDA) is a novel method of multicategory
supervised learning [\citet{lange08}].
It discriminates among categories by minimizing $\varepsilon$-insensitive
loss plus a penalty. For reasons of symmetry,
the vertices corresponding to the different classes are taken to be equidistant.
With two categories, the points $-1$ and 1 on the real line suffice for
discrimination.
With three categories there is no way of choosing three equidistant
points on the line.
Therefore, we pass to the plane and choose the vertices of an
equilateral triangle. In general with
$k>3$ categories, we choose the vertices $v_1,\ldots,v_k$ of a regular
simplex in $\mathbb{R}^{k-1}$.
Among the many ways of constructing a regular simplex, we prefer the
simple definition
\begin{equation}\label{vertices}
v_j =\cases{
 (k-1)^{-1/2}{\mathbf{1}}, &\quad if  $j=1$, \cr
 c\mathbf{1}+de_{j-1}, &\quad if $2 \leq j \leq k$,}
\end{equation}
where
\[
c = -\frac{1+\sqrt{k}}{(k-1)^{3/2}}, \qquad d = \sqrt{\frac{k}{k-1}} ,
\]
and $e_j$ is the $j$th coordinate vector in $\mathbb{R}^{k-1}$. This
puts the vertices on the surface of the unit ball
in $\mathbb{R}^{k-1}$. It is impossible to situate more than $k$
equidistant points in $\mathbb{R}^{k-1}$.

Suppose $Y$ and $X$ denote the class indicator and feature vector of a
random case. The vector $Y$ coincides with one of the vertices of the
simplex. Given a loss function $L(y,x)$, discriminant analysis seeks to
minimize the expected loss
\[
\mathrm{E}[L(Y,X)] = \mathrm{E}\{\mathrm{E}[L(Y,X)\vert X]\}.
\]
This is achieved empirically by minimizing the average conditional loss
$n^{-1}\times\sum_{i=1}^n\!L(y_i,x_i)$. To maintain parsimony, VDA postulates
the linear regression model $y=A x+b$, where $A=(a_{jl})$ is a $(k-1)
\times p$ matrix of slopes and $b=(b_j)$ is a $k-1$ column vector of
intercepts. Overfitting is avoided by imposing penalties on the slopes
$a_{jl}$ but not on the intercepts $b_j$. In VDA we take the loss
function for case $i$ to be $g(y_i-Ax_i-b)$, where $g(z)$ is the
$\varepsilon$-insensitive Euclidean distance
\[
g(z)= \|z\|_{2,\varepsilon}= \max\{\|z\|_2-\varepsilon,0\} .
\]
Classification proceeds by minimizing the objective function
\begin{equation}\label{discriminant_objective}
f(\theta) = \frac{1}{n} \sum_{i=1}^n g(y_i-A x_i -b) + \lambda P(A),
\end{equation}
where $\theta=(A,b)$, and $P(A)$ is the penalty on the matrix of slopes
$A$. Since the loss function is convex, it is clearly advantageous to
take $P(A)$ to be convex as well. In VDA$_{\mathrm{R}}$ the ridge penalty
$P(A) = \sum_j \sum_l a_{jl}^2$ is employed. Because of its near strict
convexity, the objective function $f(\theta)$ usually has a unique
minimum. Once $A$ and $b$ are estimated, we can assign a new case to
the closest vertex, and hence category.

For prediction purposes, VDA$_{\mathrm{R}}$ is competitive in statistical
accuracy and computational speed with the best available algorithms for
discriminant analysis [\citet{lange08}]. Unfortunately, it suffers two
limitations. First, although it shrinks estimates toward 0, it is
incapable of model selection unless one imposes an arbitrary cutoff on
parameter magnitudes. Second, its computational complexity scales as
$O(p^3)$ for $p$ predictors. This barrier puts problems with ten of
thousands of predictors beyond its reach. Modern genomics problems
involve hundreds of thousands to millions of predictors. The twin
predicaments of model selection and computational complexity have
prompted us to redesign VDA with different penalties and a different
optimization algorithm.

\subsection{A toy example for vertex discriminant analysis}

The use of $\varepsilon$-insen\-si\-tive loss is based on the assumption that
it makes little difference how close a linear predictor is to its class
indicator when an observation is correctly classified. Here $\varepsilon$
is the radius of the circle/ball around each vertex. Training
observations on the boundary or exterior of the $\varepsilon$-insensitive
balls act as support vectors and exhibit sensitivity. Observations
falling within an $\varepsilon$-insensitive ball exhibit insensitivity and
do not directly contribute to the estimation of regression
coefficients. The definition of $\varepsilon$-insensitive loss through
Euclidean distance rather than squared Euclidean distance makes
classification more resistant to outliers. The following small
simulation example demonstrates the importance of creating the dead
zones where observations receive a loss of 0. These zones render
estimation and classification highly nonlinear.

We generated 300 training observations equally distributed over $k=3$
classes. To each observation $i$ we attached a normally distributed
predictor $x_i$ with variance $1$ and mean
\[
\mu =\cases{-4, &\quad \mbox{class} $= 1$, \cr
            0, &\quad \mbox{class} $= 2$, \cr
            4, &\quad \mbox{class} $= 3$.}
\]
We then compared four methods: (a) least squares with class indicators
$v_j$ equated to the standard unit vectors $e_j$ in $\mathsf{R}^3$
(indicator regression); (b) least squares with class indicators $v_j$
equated to the vertices of an equilateral triangle inscribed on the
unit circle as described in (\ref{vertices}); (c) $\varepsilon
$-insensitive loss with the triangular vertices and $\varepsilon=0.6$; and
(d) $\varepsilon$-insensitive loss with the triangular vertices and
$\varepsilon=1/2\sqrt{2k/(k-1)}=0.866$. Because there is only a single
predictor, all four methods omit penalization. As advocated in \citet
{lange08}, method (d) adopts the maximum value of $\varepsilon$ consistent
with nonoverlapping balls.

\begin{figure}

\includegraphics{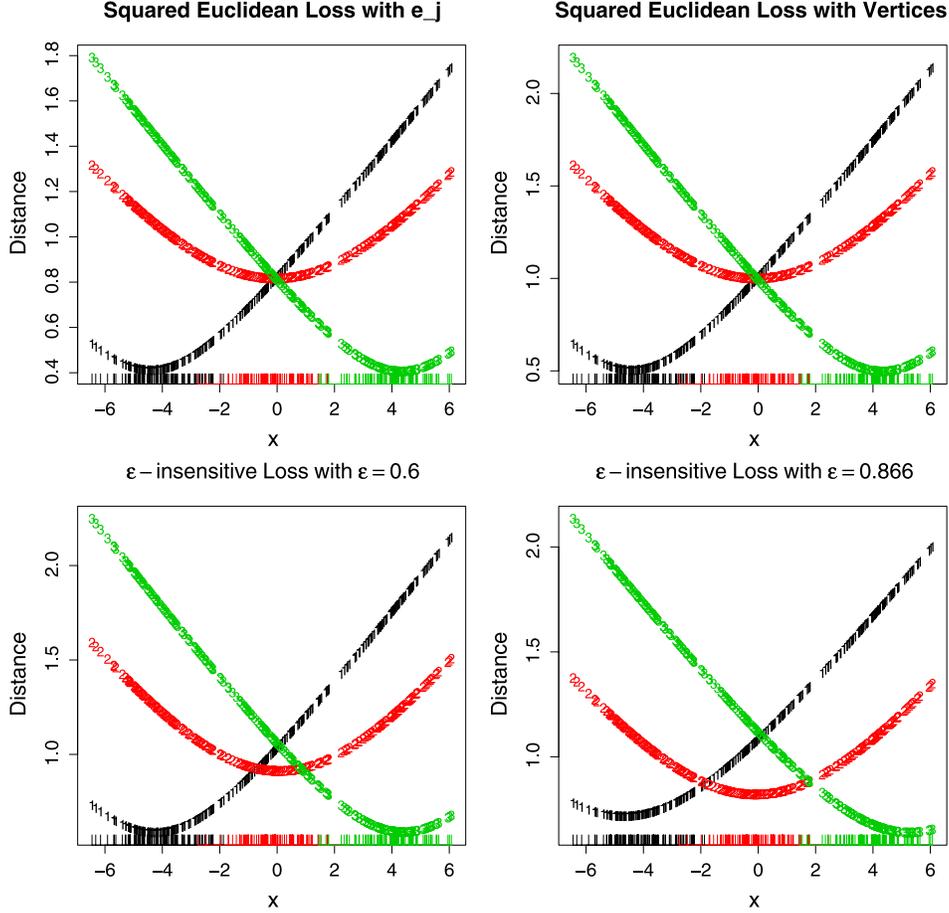}

\caption{Distance to class indicators. The upper left plot shows
observed distances under squared Euclidean loss with class indicators
$v_j$ equated to the standard unit vectors in $\mathsf{R}^3$. The upper
right plot shows observed distances under squared Euclidean loss with
class indicators equated to the vertices of an equilateral triangle.
The lower left plot shows observed distances under $\varepsilon
$-insensitive loss with the triangular vertices and $\varepsilon=0.6$.
Finally, the lower right plot shows observed distances under $\varepsilon
$-insensitive loss with the triangular vertices and $\varepsilon=1/2\sqrt
{2k/(k-1)}=0.866$. In the lower right plot, it is clear that
observations with $x<-2$ will be predicted as class 1 (black),
observations with $x>2$ will be predicted as class 3 (green), and
observations with $-2 \leq x \leq2$ will be predicted as class 2
(red). This is consistent with the true classes shown on the $x$-axis.}\label{plot_test}
\end{figure}

Figure \ref{plot_test} plots the three distances $x_i \rightarrow\|
{\hat y}_i-v_j\|$ between the predicted value
$\hat{y}_i$ for observation $i$ and each of the three vertices $v_j$.
An observation is assigned to the class whose vertex is closest. It is
evident from these plots that squared Euclidean loss fails to identify
class 2, which is dominated and masked by the other two classes (upper
two panels of Figure \ref{plot_test}). With surrounding balls of small
radius, class 2 can be identified but the misclassification rate is
high (13\%, lower left plot). With surrounding balls of the maximum
legal radius, $\varepsilon$-insensitive loss readily distinguishes all
three classes with a low misclassification rate ($2.67\%$, lower right
plot). This example nicely illustrates the importance of the dead zones
integral to $\varepsilon$-insensitive loss. Our previous paper
[\citet{lange08}]
reaches essentially the same conclusions by posing
discrimination with three classes as a problem in one-dimensional
regression. Section \ref{Fisher_consistency_section} discusses how
$\varepsilon$-insensitive loss achieves Fisher consistency. The dead zones
figure prominently in the derivation of consistency.

In these four examples masking is neutralized. Because our proof of
Fisher consistency requires nonlinear as well as linear functions, the
possibility of masking still exists in practice. Inclusion of nonlinear
combinations of predictors, say products of predictors, may remedy the
situation. Of course, creating extra predictors highlights the need for
rigorous model selection and fast computation.

\subsection{Modified $\varepsilon$-insensitive loss}

The kinks in $\varepsilon$-insensitive loss have the potential to make
Newton's method behave erratically in cyclic coordinate descent. It is
possible to avoid this pitfall by substituting a similar loss function
that is smoother and still preserves convexity. Suppose $f(s)$ is an
increasing convex function defined on $[0,\infty)$. If $\|x\|$ denotes
the Euclidean norm of $x$, then the function $f(\|x\|)$ is convex. This
fact follows from the inequalities
\begin{eqnarray*}
f[\|\alpha x + (1-\alpha)y\|] &\le& f[\alpha\|x\|+(1-\alpha)\|y\|]\\
 &\le& \alpha f(\|x\|)+(1-\alpha)f(\|y\|)
\end{eqnarray*}
for $\alpha\in[0,1]$. It seems reasonable to perturb the $\varepsilon$-insensitive
 function as little as possible. This suggests eliminating
the corner near $s=\varepsilon$. Thus, we define $f(s)$ to be 0 on the
interval $[0,\varepsilon-\delta)$, a polynomial on the interval
$[\varepsilon-\delta,\varepsilon+\delta]$, and $s-\varepsilon$ on the interval
$(\varepsilon+\delta,\infty)$. Here we obviously require $0 < \delta< \varepsilon$.

There are two good candidate polynomials. The first is the quadratic
\[
p_2(s) = \frac{(s-\varepsilon+\delta)^2}{4\delta}.
\]
This function matches the function values and the first derivatives of
the two linear pieces at the join points $\varepsilon-\delta$ and
$\varepsilon
+\delta$. Indeed, brief calculations show that
\[
p_2(\varepsilon-\delta)= 0,\qquad p_2'(\varepsilon-\delta)=0,\qquad p_2(\varepsilon+\delta)= \delta,\qquad p_2'(\varepsilon+\delta)= 1.
\]
Unfortunately, the second derivative $p_2''(s) = (2 \delta)^{-1}$ does
not match the vanishing second derivatives of the two linear pieces at
the join points. Clearly, $p_2(s)$ is increasing and convex on the open
interval $(\varepsilon-\delta,\varepsilon+\delta)$.

A more complicated choice is the quartic polynomial
\[
p_4(s) = \frac{(s-\varepsilon+\delta)^3(3 \delta-s+\varepsilon)}{16\delta^3}.
\]
Now we have
\begin{eqnarray*}
p_4(\varepsilon-\delta) &=&  0,\qquad p_4'(\varepsilon-\delta) =0,\qquad p_4''(\varepsilon-\delta) = 0, \\
p_4(\varepsilon+\delta) &=&  \delta,\qquad p_4'(\varepsilon+\delta) = 1,\qquad p_4''(\varepsilon+\delta)= 0.
\end{eqnarray*}
Both the first and second derivatives
\begin{eqnarray*}
p_4'(s) &=&\frac{(s-\varepsilon+\delta)^2(2\delta-s+\varepsilon)}{4\delta^3}, \\
p_4''(s) &=& \frac{3(s-\varepsilon+\delta)(\varepsilon+\delta-s)}{4\delta^3}
\end{eqnarray*}
are positive throughout the open interval $(\varepsilon-\delta,\varepsilon+\delta)$.
 The second derivative attains its maximum value of $\frac{3}{4\delta}$ at the midpoint $\varepsilon$. Thus, $p_4(s)$ is increasing and
convex on the same interval.
We now write $p(s)$ for the function equal to 0 on $[0,\varepsilon-\delta)$, to $p_4(s)$ on
$[\varepsilon-\delta,\varepsilon+\delta]$, and to $s-\varepsilon$ on
$(\varepsilon+\delta,\infty)$.

\subsection{Cyclic coordinate descent}

In our modified VDA the alternative loss function $p(\|y_i-Ax_i-b\|)$
is twice continuously differentiable. This has the advantage of
allowing us to implement Newton's method. If we abbreviate $y_i-Ax_i-b$
by $r_i$, then applying the chain rule repeatedly yields the partial derivatives
\begin{eqnarray*}
\frac{\partial}{\partial b_j}p(\|r_i\|) & = & - \frac{p'(\|r_i\|)}{\|r_i\|} r_{ij}, \\
\frac{\partial^2}{\partial b_j^2}p(\|r_i\|) & = & \frac{p''(\|r_i\|)}{\|r_i\|^2} r_{ij}^2+\frac{p'(\|r_i\|)}{\|r_i\|}
\biggl(1-\frac{r_{ij}^2}{\|r_i\|^2}\biggr), \\
\frac{\partial}{\partial a_{jl}}p(\|r_i\|) & = & - \frac{p'(\|r_i\|)}{\|r_i\|} r_{ij}x_{il}, \\
\frac{\partial^2}{\partial a_{jl}^2}p(\|r_i\|) & = & \frac{p''(\|r_i\|)}{\|r_i\|^2} (r_{ij}x_{il})^2+\frac{p'(\|r_i\|)}{\|r_i\|}
\biggl[x_{il}^2-\frac{(r_{ij}x_{il})^2}{\|r_i\|^2}\biggr].
\end{eqnarray*}
The only vector operation required to form these partial derivatives is
computation of the norm $\|r_i\|$. As long as the number of categories
is small and we update the residuals $r_i$ as we go, the norms are
quick to compute.

Our overall objective function $f(\theta)$ is given in (\ref{discriminant_objective}) with
%
\begin{equation}\label{loss_modified}
\qquad g(v) = \cases{
\displaystyle{\|v\|_2-\varepsilon}, & \quad if $\|v\|_2 > \varepsilon+\delta$, \cr
\displaystyle{\frac{(\|v\|_2-\varepsilon+\delta)^3(3 \delta-\|v\|_2+\varepsilon)}{16\delta^3}}, & \quad if $\|v\|_2 \in[\varepsilon-\delta,\varepsilon+\delta]$,\cr
0, & \quad if $\|v\|_2 < \varepsilon-\delta$,}
\end{equation}
replacing the $\varepsilon$-insensitive loss $g(v)=\|z\|_{2,\varepsilon}$
throughout. To minimize this objective function in the presence of a
large number of predictors, we use the cyclic version of coordinate
descent highlighted by \citet{friedman07} and
\citet{wu08}. Cyclic coordinate descent avoids the bottlenecks of
ordinary regression, namely matrix diagonalization, matrix inversion,
and the solution of large systems of linear equations. It is usually
fast and always numerically stable for smooth convex objective functions.

Consider now the convex lasso penalty $P(A)=\sum_{j=1}^{k-1}\sum_{l=1}^p |a_{jl}|$.
 Although the objective function
$f(\theta)$ is nondifferentiable, it possesses forward and backward
directional derivatives along each coordinate direction. If $e_{jl}$ is
the coordinate direction along which $a_{jl}$ varies, then the forward
and backward directional derivatives are
\begin{eqnarray*}
d_{e_{jl}}f(\theta) & = & \lim_{\tau\downarrow 0} \frac{f(\theta+\tau
e_{jl}) - f(\theta)}{\tau} \\
& = & \frac{1}{n} \sum_{i=1}^n \frac{\partial}{\partial a_{jl}} g(r_i)
+ (-1)^{I(a_{jl} < 0)}\lambda
\end{eqnarray*}
and
\begin{eqnarray*}
d_{-e_{jl}}f(\theta) & = & \lim_{\tau\downarrow 0} \frac{f(\theta
-\tau e_{jl}) - f(\theta)}{\tau} \\
& = & -\frac{1}{n} \sum_{i=1}^n \frac{\partial}{\partial a_{jl}} g(r_i)
+ (-1)^{I(a_{jl} > 0)}\lambda,
\end{eqnarray*}
where $I(\cdot)$ is an indicator function taking value 1 if the
condition in the parentheses is satisfied and 0 otherwise.

Newton's method for updating a single intercept parameter of $f(\theta
)$ works well because there is no lasso penalty. For a slope parameter
$a_{jl}$, the lasso penalty intervenes, and we must take particular
care at the origin. If both of the directional derivatives
$d_{e_{jl}}f(\theta)$ and $d_{-e_{jl}}f(\theta)$ are nonnegative, then
the origin furnishes the minimum of the objective function along the
$e_{jl}$ coordinate. If either directional derivative is negative, then
we must solve for the minimum in the corresponding direction. Both
directional derivatives cannot be negative because this contradicts the
convexity of $f(\theta)$. In practice, we start all parameters at the
origin. For underdetermined problems with just a few relevant
predictors, most updates are skipped, and many parameters never budge
from their starting values of 0. This simple fact plus the complete
absence of matrix operations explains the speed of cyclic coordinate
descent. It inherits its numerical stability from the descent property
of each update.

Newton's method for updating $a_{jl}$ iterates according to
\[
a_{jl}^{m+1}  =  a_{jl}^m - \frac{(1/n) \sum_{i=1}^n
\frac{\partial}{\partial a_{jl}} g(r_i^m) + (-1)^{I(a_{jl}^m < 0)}\lambda}
{(1/n) \sum_{i=1}^n \frac{\partial^2}{\partial a_{jl}^2}g(r_i^m)},
\]
where $r_i^m$ is the value of the $i$th residual at iteration $m$. In
general, one should check that the objective function is driven
downhill. If the descent property fails, then the simple remedy of step
halving is available. The Newton update for an intercept is
\[
b_j^{m+1}  =  b_j^m - \frac{(1/n) \sum_{i=1}^n \frac{\partial}{\partial b_j} g(r_i^m)}
{(1/n)\sum_{i=1}^n \frac{\partial^2}{\partial b_j^2} g(r_i^m)}.
\]
%

\section{Penalty for grouped effects}\label{sec3}

\subsection{Euclidean penalty}

In model selection it is often desirable to impose coordinated
penalties that include or exclude all of the parameters in a group. In
multicategory classification, the slopes of a single predictor for
different dimensions of $\mathbb{R}^{k-1}$ form a natural group. In
other words, the parameter group for predictor $l$ is the $l$th column
$a_l=(a_{1l},\dots,a_{k-1,l})^t$ of the slope matrix $A$. The lasso
penalty $\lambda_L \|a_l\|_1$ and the ridge penalty $\lambda_R \|a_l\|
_2^2$ separate parameters and do not qualify as sensible group
penalties. The scaled Euclidean norm $\lambda_E \|a_l\|_2$ is an ideal
group penalty since it couples parameters and preserves convexity
[\citet{wu08}; \citet{wuzou08}].

The Euclidean penalty possesses several other desirable features.
First, it reduces to a lasso penalty $\lambda|a_{jl}|$ on $a_{jl}$
whenever $a_{ml}=0$ for $m \ne j$. This feature of the penalty enforces
parsimony in model selection. Second, the Euclidean penalty is
continuously differentiable in $a_{l}$ whenever $a_{l}$ is nontrivial.
Third, the Euclidean penalty is spherically symmetric. This makes the
specific orientation of the simplex irrelevant. If one applies an
orthogonal transformation $O$ to the simplex, then the transformed
vertices $Oy$ are still equidistant. Furthermore, the new and old
versions of the objective functions satisfy
\begin{eqnarray*}
&&\frac{1}{n} \sum_{i=1}^n g(y_i-Ax_i-b)+\lambda_E \sum_{l=1}^p \|
a_l\|_2 \\
&&\qquad =\frac{1}{n} \sum_{i=1}^n g(Oy_i-O Ax_i-O b)+\lambda_E \sum
_{l=1}^p \|O a_l\| .
\end{eqnarray*}
Thus, any minimum of one orientation is easily transformed into a
minimum of the other, and predictors active under one orientation are
active under the other orientation. For instance, if the estimates for
the original objective function are $\hat{A}$ and $\hat{b}$, then the
estimates for the transformed objective
function are $O \hat{A}$ and $O \hat{b}$.

\subsection{Coordinate descent under a Euclidean penalty}

In modified VDA with grouped effects, we minimize the objective function
%
\begin{equation}\label{loss_group}
f(\theta) =  \sum_{i=1}^n g(y_i-Ax_i-b) + \lambda_L \sum_{j=1}^{k-1}\sum_{l=1}^p |a_{jl}|
+ \lambda_E \sum_{l=1}^p \|a_l \|_2.
\end{equation}
If the tuning parameter for the Euclidean penalty $\lambda_E=0$, then
the penalty reduces to the lasso. On the other hand, when the tuning
parameter for the lasso penalty $\lambda_L=0$, only group penalties
enter the picture. Mixed penalties with $\lambda_L>0$ and $\lambda_E>0$
enforce shrinkage in both ways. All mixed penalties are norms on $A$
and therefore convex functions.

The partial derivatives of the Euclidean penalty are similar to those
of the loss function $g(v)$. There are two cases to consider. If $\|
a_l\| = 0$, then the forward and backward derivatives of $\lambda_E \|
a_l\|$ with respect to $a_{jl}$ are both $\lambda_E$. The forward and
backward second derivatives vanish. If $\|a_l\|>0$, then $\|a_l\|$ is
differentiable and
\begin{eqnarray*}
\frac{\partial}{\partial a_{jl}}\lambda_E \|a_l\| &=& \lambda_E
\frac{a_{jl}}{\|a_l\|},\\
 \frac{\partial^2}{\partial a_{jl}^2} \lambda_E \|a_l\| &=&
\frac{\lambda_E}{\|a_l\|}\biggl(1-\frac{a_{jl}^2}{r\|a_l\|^2}\biggr).
\end{eqnarray*}

\section{Fisher consistency of $\varepsilon$-insensitive loss}\label{Fisher_consistency_section}

A loss function $L(y,x)$ is Fisher consistent if minimizing its average
risk $\mathrm{E}\{L[f(X),Y)]\}$ leads to the Bayes optimal decision rule.
Fisher consistency is about the least one can ask of a loss function.
Our previous development of VDA omits this crucial topic, so we take it
up now for $\varepsilon$-insensitive loss without a penalty. The empirical
loss minimized in VDA is
\[
\mathrm{EML}_n(L,f) = \frac{1}{n} \sum_{i=1}^n L[f(x_i),y_i]= \frac{1}{n} \sum_{i=1}^n \|y_i-f(x_i)\|_{\varepsilon},
\]
and the VDA classifier is obtained by solving
\[
\hat{f} = \operatorname{arg}\min_{f \in \mathcal{F}_n} \mathrm{EML}_n(L,f),
\]
where $\mathcal{F}_n$ is the space of linear functions in the predictor
matrix $(x_{ij})$. This space is determined by the slope matrix $A$ and
the intercept vector $b$. Once these are estimated, we assign a new
case to the class attaining
%
\begin{equation}\label{classifier}
\operatorname{arg\,min}\limits_{j \in\{1,\ldots, k\}} \|v_j - \hat{f}\| =
\operatorname{arg\,min}\limits_{j \in\{1,\ldots, k\}} \|v_j - \hat{A}x - \hat{b}\|.
\end{equation}
If we define $p_j(x) = \Pr(Y =v_j \vert X=x)$ to be the conditional
probability of class $j$ given feature vector $x$, then Fisher
consistency demands that the minimizer $f^*(x)=\operatorname{arg\,min}\mathrm{E}[\|Y-f(X)\|
_{\varepsilon} \mid X=x]$ satisfy
\[
\operatorname{arg\,min}\limits_{j \in\{1,\ldots, k\}} \|v_j - f^*(x) \| = \operatorname{arg\,max}\limits_{j\in\{1,\ldots, k\}} p_j(x).
\]
Here distance is ordinary Euclidean distance, and $f^*(x)$ is not
constrained to be linear in $x$. In the \ref{suppA} [\citet{wu10supplement}] we prove the following proposition.

\begin{proposition} \label{proposition1}
If a minimizer $f^*(x)$ of $\mathrm{E}[\|Y-f(X)\|_{\varepsilon} \mid X=x]$ with
$\varepsilon=\frac{1}{2}\sqrt{2k/(k-1)}$ lies closest to vertex $v_l$,
then $p_l(x) = \max_j p_j(x)$. Either $f^*(x)$ occurs exterior to all
of the $\varepsilon$-insensitive balls or on the boundary of the ball
surrounding $v_l$. The assigned vertex $v_l$ is unique if the $p_j(x)$
are distinct.
\end{proposition}

\begin{figure}

\includegraphics{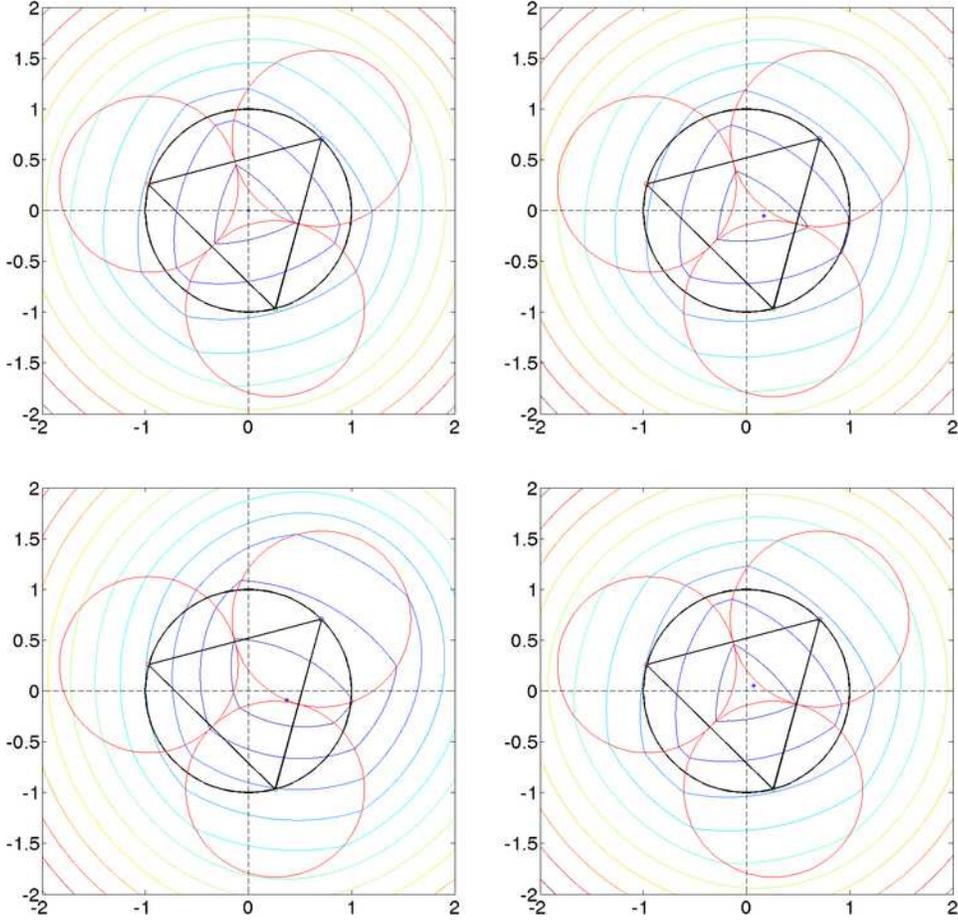}

\caption{Contour plots of $\sum_j p_j \|v_j-z\|_\varepsilon$ with $k=3$
and $\varepsilon=\frac{1}{2}\sqrt{k/(k-1)}$ for different $p$'s. Upper
left panel $p=(1/3,1/3,1/3)$, upper right panel $p=(0.37,0.37,0.26)$,
lower left panel $p=(0.6,0.3,0.1)$, and lower right panel
$p=(1/3+t,1/3-0.25t,1/3-0.75t)$ with $t=0.025$.} \label{fig_loss}
\end{figure}

To help the reader better understand the behavior of the nonlinear
function $z \mapsto\sum_j p_j \|v_j-z\|_\varepsilon$, we plot it and its
contour lines in Figure \ref{fig_loss} for $k=3$ classes. The three
class vertices are labeled clockwise starting with vertex 1 in the
first quadrant. Here we take $\varepsilon=\frac{1}{2}\sqrt{2k/(k-1)}$ to
be the largest possible value avoiding overlap of the interiors of the
$\varepsilon$-insensitive balls around each vertex of the regular simplex.
Figure \ref{fig_loss} demonstrates that the optimal point varies with
the probability vector $p$. When the highest probabilities are not
unique (upper two panels of Figure \ref{fig_loss}), the optimal point
falls symmetrically between the competing vertices. When there is a
dominant class (lower left panel of Figure \ref{fig_loss}), the optimal
point falls on the boundary of the dominant ball. In the first case
with $p=(1/3,1/3,1/3)$, if we slowly increase $p_1$ and decrease $p_2$
and $p_3$ symmetrically, then the optimal point moves from the origin
to the boundary of the ball surrounding vertex 1 (lower right of Figure
\ref{fig_loss}).

\section{Stability selection}\label{sec5}

Stability selection [Meinshausen and Buehlmann\break(\citeyear{meinshausen09})] involves subsampling the data
and keeping a tally of how often a given variable is selected. Each new
subsample represents a random choice of half of the existing cases. Let
$\hat{\Pi}_k^\lambda$ be the empirical probability over the subsamples
that variable $k$ is selected under a particular value $\lambda$ of the
penalty tuning constant; the universe of relevant tuning constants is
denoted by $\Lambda$. \citet{meinshausen09} recommend 100 subsamples;
the choice of $\Lambda$ is left to the discretion of the user. A
predictor $k$ is considered pertinent whenever $\max_{\lambda\in
\Lambda} \hat{\Pi}_k^\lambda\ge\pi$ for some fixed threshold
$\pi>\frac{1}{2}$. The set of pertinent predictors $\hat{S}^\mathrm{stable}$ is
the final (or stable) set of predictors determined by this criterion.

One of the appealing features of stability selection is that it
controls for the number of false positives. Under certain natural
assumptions, \citet{meinshausen09} demonstrate that the expected number
of false positives among the stable set is bounded above by the
constant $q^2/[(2\pi-1)p]$, where $q$ is the average size of the random
union $\hat{S}^\Lambda= \bigcup_{\lambda\in\Lambda} \hat{S}^\lambda$,
and $\hat{S}^\lambda$ is the set of predictors selected at the given
penalty level $\lambda$ in the corresponding random subsample.

\section{Simulation examples}\label{sec6}

\subsection{Simulation example 1}

The simulation examples of \citet{wang07} offer an opportunity to
compare VDA$_{\mathrm{LE}}$, VDA$_{\mathrm L}$, and VDA$_{\mathrm E}$ to the highly
effective methods OVR and L1MSVM [\citet{wang07}]. Example 1 of \citet{wang07}
specifies either $k=4$ and $k=8$ classes, $n=20k$ training
observations, and $p=100$ predictors. When observation $i$ belongs to class
$c \in\{1,\dots,k\}$, its $j$th predictor $x_{ij}$ is defined by
\[
x_{ij}  = \cases{
u_{ij}+a_j, &\quad if $j = 1,2$, \cr
u_{ij}, & \quad if $j = 3,\dots,100$,}
\]
where $a_1=d \cdot\cos[2(c-1)\pi/k]$ and $a_2=d \cdot\sin
[2(c-1)\pi
/k]$. The $u_{ij}$ are independent normal variates with mean 0 and
variance 1. Only the first and second predictors $x_1$ and $x_2$ depend
on an observation's class. The constant $d$ determines the degree of
overlap of the classes. The various combinations of $k \in\{4,8\}$ and
$d \in\{1,2,3\}$ generate six datasets of varying difficulty. The
three datasets with $k=4$ are underdetermined since $n=80<p=100$; the
datasets with $k=8$ are overdetermined since $n=160>p=100$. As
recommended in \citet{lange08}, we take $\varepsilon=\frac{1}{2} \sqrt
{2k/(k-1)}$, the largest possible value avoiding overlap of the
interiors of the $\varepsilon$-insensitive balls around the vertices of
the regular simplex. For all five methods, we chose the penalty tuning
constants by minimizing assignment error on a separate testing sample
with 20,000 observations. Table~\ref{sim_wang1} reports Bayes errors,
optimal testing errors, number of variables selected, and elapsed
training time in seconds ($\times10^4$) averaged over 100 random replicates.

Table \ref{sim_wang1} shows that VDA$_{\mathrm{LE}}$ and VDA$_{\mathrm E}$
outperform VDA$_{\mathrm L}$ across all six datasets. Testing error is
closely tied to the number of predictors selected. The addition of a
lasso penalty gives VDA$_{\mathrm{LE}}$ a slight edge over VDA$_{\mathrm E}$.
The competing method L1MSVM performs best overall by a narrow margin.
The true predictors $x_1$ and $x_2$ are always selected by all three
VDA methods. In reporting their results for L1MSVM and OVR, \citet
{wang07} omit mentioning the number of true predictors selected and
computing times.

\begin{sidewaystable}
\tabcolsep=0pt
\tablewidth=\textheight
\tablewidth=\textwidth
\caption{Comparison of VDA$_{\mathrm {LE}}$, VDA$_{\mathrm L}$, VDA$_{\mathrm E}$,
L1MSVM, and OVR on simulation example
1. Here $p=100$ and $n=20k$; $d$ is the multiplier in each simulation.
Column 3 lists the Bayes error as a percentage. Column 5 reports the
mean and 10\%, 50\%, and 90\% percentiles of the number of nonzero
variables. The remaining columns report average testing error, average
number of nonzero variables, and average training time in seconds
($\times10^4$) over 100 random replicates. The corrresponding standard
errors for these averages appear in parentheses. Tuning constants were
chosen to minimize error over a larger independent dataset with 20,000
observations. The results of L1MSVM and OVR are taken from the paper of
Wang and Shen (\protect\citeyear{wang07})}\label{sim_wang1}
\begin{tabular*}{\textwidth}{@{\extracolsep{\fill}}lcd{2.2}d{2.7}d{3.16}f{3.3}d{2.7}d{1.7}d{2.7}@{\hspace*{2pt}}d{2.8}@{}}
\hline
&  &  & \multicolumn{3}{c}{\textbf{VDA}$_{\mathbf{LE}}$\textbf{,} \textbf{VDA}$_{\mathbf{L}}$
\textbf{and} \textbf{VDA}$_{\mathbf{E}}$} & \multicolumn{2}{c}{\textbf{L1MSVM}} &
\multicolumn{2}{c@{}}{\textbf{OVR}}\\[-5pt]
& & &\multicolumn{3}{c}{\hrulefill} &\multicolumn{2}{c}{\hrulefill}
&\multicolumn{2}{c@{}}{\hrulefill}\\
$\bolds k$ & \multicolumn{1}{c}{$\bolds d$} & \multicolumn{1}{c}{\textbf{Bayes (\%)}} & \multicolumn{1}{c}{\textbf{Error (\%)}}
& \multicolumn{1}{c}{\textbf{\# Var}}
&\multicolumn{1}{c}{\textbf{Time}} &\multicolumn{1}{c}{\textbf{Error (\%)}} & \multicolumn{1}{c}{\textbf{\# Var}}
& \multicolumn{1}{c}{\textbf{Error (\%)}} & \multicolumn{1}{c@{}}{\textbf{\# Var}}\\
\hline
4 & 1 & 36.42 & 43.19\ (0.09) & 2.80\ (0.11)\ 2,2,4 & 80,(3) & 42.20\ (0.09) &
2.20\ (0.05) & 56.87\ (0.25) & 67.17\ (1.93) \\
& & & 44.95\ (0.22) & 8.16\ (0.84)\ 2,5,16 & 73,(3) \\
& & & 43.27\ (0.08) & 2.42\ (0.05)\ 2,2,3 & 114,(3) \\
{} & 2 & 14.47 & 15.31\ (0.03) & 2.07\ (0.02)\ 2,2,2 & 115,(4) & 15.18\ (0.04)
& 2.06\ (0.02) & 16.21\ (0.09) & 5.72\ (0.38) \\
& & & 16.22\ (0.11) & 3.79\ (0.28)\ 2,3,8 & 112,(3) \\
& & & 15.54\ (0.04) & 2.13\ (0.04)\ 2,2,3 & 139,(3) \\
 & 3 & 3.33 & 3.40\ (0.01) & \multicolumn{1}{e{9.6}}{2\ (0)\ 2,2,2} & 182,(13) & 3.35\ (0.02) &
2.02\ (0.01) & 3.50\ (0.02) & 2.51\ (0.13) \\
& & & 3.80\ (0.04) & 3.18\ (0.18)\ 2,2,5 & 145,(7) \\
& & & 3.52\ (0.01) & 2.12\ (0.04)\ 2,2,2 & 197,(8) \\
8 & 1 & 64.85 & 70.94\ (0.11) & 2.43\ (0.08)\ 2,2,4 & 312,(10) & 70.47\ (0.10)
& 3.51\ (0.16) & 79.76\ (0.07) & 98.18\ (0.29) \\
& & & 74.77\ (0.09) & 23.19\ (1.99)\ 2,18,51 & 278,(6) \\
& & & 70.81\ (0.10) & 2.57\ (0.09)\ 2,2,4 & 387,(3) \\
 & 2 & 43.82 & 51.09\ (0.24) & 2.27\ (0.06)\ 2,2,3 & 351,(10) & 46.86\ (0.12)
& 3.02\ (0.12) & 66.72\ (0.11) & 95.43\ (0.25) \\
& & & 58.37\ (0.11) & 33.34\ (1.48)\ 15,32,52 & 269,(6) \\
& & & 50.50\ (0.22) & 2.17\ (0.05)\ 2,2,3 & 355,(9) \\
 & 3 & 25.06 & 37.93\ (0.40) & 2.23\ (0.05)\ 2,2,3 & 436,(9) & 27.95\ (0.13)
& 2.75\ (0.17) & 55.84\ (0.12) & 93.37\ (0.21) \\
& & & 46.91\ (0.15) & 33.88\ (1.30)\ 17,32,50 & 264,(5) \\
& & & 33.26\ (0.36) & 2.02\ (0.01)\ 2,2,2 & 462,(4) \\
\hline
\end{tabular*}
\end{sidewaystable}

\subsection{Simulation example 2}

In the second example of \citet{wang07}, $k=3$, $n=60$, and $p$ varies
over the set $\{10,20,40,80,160\}$. We now compare the three modified
VDA methods with L1MSVM [\citet{wang07}] and L2MSVM [\citet{lee04}]. The 60
training cases are spread evenly across the three classes. For $p$
equal to 10, 20, and 40, discriminant analysis is overdetermined; the
reverse holds for $p$ equal to 80 and 160. The predictors $x_{ij}$ are
independent normal deviates with variance 1 and mean 0 for $j>2$. For
$j \le2$, $x_{ij}$ have mean $a_j$ with
\[
(a_1,a_2) = \cases{
\bigl(\sqrt{2}, \sqrt{2}\bigr), & \quad for class $1$, \cr
\bigl(-\sqrt{2}, -\sqrt{2}\bigr), & \quad  for class $2$, \cr
\bigl(\sqrt{2}, -\sqrt{2}\bigr), & \quad for class $3$.}
\]
Only the first two predictors are relevant to classification.

We computed VDA testing errors over an independent dataset with 30,000
observations and chose penalty tuning constants to minimize testing
error over a grid of values. Table \ref{sim_wang2} reports Bayes
errors, optimal testing errors, number of variables selected, and
training times in seconds ($\times10^4$) averaged across 100 random
replicates. In this example VDA$_{\mathrm {LE}}$, VDA$_{\mathrm {E}}$, and L1MSVM
rank first, second, and third, respectively, in testing error. Again
there is a strong correlation between testing error and number of
predictors selected, and the lasso penalty is effective in combination
with the Euclidean penalty. The true predictors $X_1$ and $X_2$ are
always selected by the three VDA methods.

\begin{table}
\tabcolsep=0 pt
\caption{Comparison of VDA$_{\mathrm {LE}}$, VDA$_{\mathrm L}$, VDA$_{\mathrm E}$,
L1MSVM, and L2MSVM on simulation example 2 with $k=3$ and $n=60$.
Column 2 lists the Bayes error as a percentage. Column 4 reports the
10\%, 50\%, and 90\% percentiles of the number of nonzero variables.
The remaining columns report average testing error, average number of
nonzero variables, and average training time in seconds ($\times10^4$)
over 100 random replicates. The corresponding standard errors for these
averages appear in parentheses. The partial results for L1MSVM and
L2MSVM are taken from the paper of Wang and Shen (\protect\citeyear{wang07})} \label{sim_wang2}
\begin{tabular*}{\textwidth}{@{\extracolsep{\fill}}ld{2.2}d{2.8}a{0.4}b{3.3}d{2.8}d{2.8}@{}}
\hline
 && \multicolumn{3}{c}{\textbf{VDA}$_{\mathbf{LE}}$\textbf{,} \textbf{VDA}$_{\mathbf{L}}$ \textbf{and} \textbf{VDA}$_{\mathbf{E}}$} & \multicolumn{1}{c}{\textbf{L1MSVM}}
 & \multicolumn{1}{c@{}}{\textbf{L2MSVM}}\\[-5pt]
& &\multicolumn{3}{c}{\hrulefill} &\multicolumn{1}{c}{\hrulefill} &\multicolumn{1}{c@{}}{\hrulefill}\\
$\bolds p$ & \multicolumn{1}{c}{\textbf{Bayes (\%)}}& \multicolumn{1}{c}{\textbf{Error (\%)}} & \multicolumn{1}{c}{\textbf{\# Var}} & \multicolumn{1}{c}{\textbf{Time}}
 & \multicolumn{1}{c}{\textbf{Error (\%)}} & \multicolumn{1}{c@{}}{\textbf{Error (\%)}} \\
\hline
\phantom{0}10 & 10.81 & 12.38\ (0.10) & 2,\ 3,\ 4 & 71,(8) & 13.61\ (0.12) & 15.44\ (0.17) \\
& & 14.42\ (0.14) & 2,\ 3,\ 10 & 50,(8) \\
& & 12.70\ (0.12) & 2,\ 3,\ 5 & 74,(8) \\
\phantom{0}20 & 10.81 & 12.65\ (0.11) & 2,\ 4,\ 6 & 104,(7) & 14.06\ (0.14) & 17.81\ (0.22) \\
& & 15.38\ (0.19) & 2,\ 4,\ 20 & 43,(7) \\
& & 13.08\ (0.13) & 3,\ 5,\ 7 & 130,(7) \\
\phantom{0}40 & 10.81 & 13.01\ (0.13) & 3,\ 5,\ 9 & 178,(10) & 14.94\ (0.14) & 20.01\ (0.22)
\\
& & 15.66\ (0.20) & 3,\ 5,\ 28 & 56,(7) \\
& & 13.50\ (0.13) & 4,\ 7,\ 10 & 247,(8) \\
\phantom{0}80 & 10.81 & 13.33\ (0.14) & 5,\ 8,\ 13 & 345,(15) & 15.68\ (0.15) & 21.81\ (0.14)
\\
& & 16.15\ (0.22) & 4,\ 8,\ 32 & 89,(8) \\
& & 13.99\ (0.15) & 8,\ 12,\ 17 & 440,(14) \\
160 & 10.81 & 14.02\ (0.14) & 3,\ 14,\ 19 & 647,(30) & 16.58\ (0.17) &
27.54\ (0.17) \\
& & 17.12\ (0.23) & 6,\ 12,\ 51 & 180,(8) \\
& & 15.08\ (0.19) & 14,\ 19,\ 26 & 830,(22) \\
\hline
\end{tabular*}
\end{table}

\begin{figure}

\includegraphics{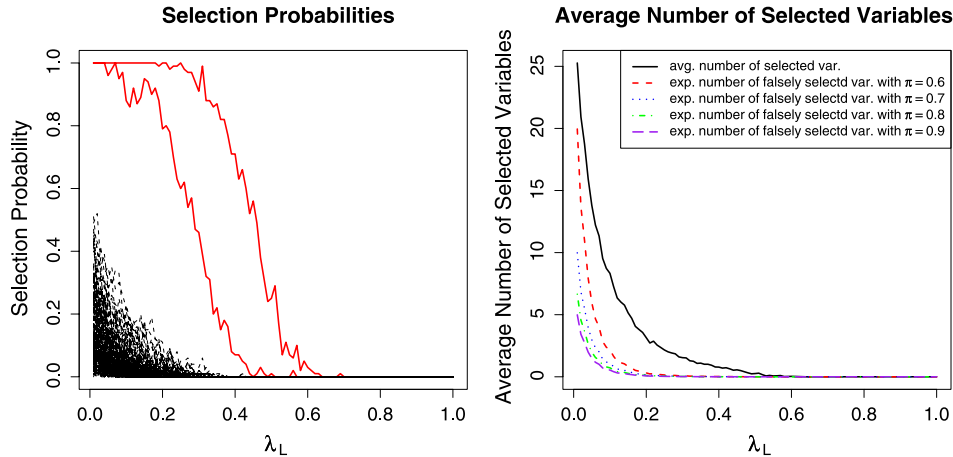}

\caption{Stability selection with VDA$_{\mathrm {LE}}$ for $p=160$. The left
panel shows the empirical selection probabilities of all 160 predictors
over 100 subsamples as a function of $\lambda_L$ for $\lambda_E$ fixed
at 0.1. The first two predictors (red solid lines) stand out from the
remaining predictors (black dash lines) with much higher selection
probabilities. The right panel plots the average number of selected
predictors (black solid line) and the expected number of falsely
selected predictors for different values of the cutoff $\pi$.}\label{fig_stability}
\end{figure}

In one of the example 2 simulations, we applied stability selection
[\citet{meinshausen09}] to eliminate false positives. The left panel of
Figure \ref{fig_stability} shows that the true predictors $X_1$ and
$X_2$ have much higher selection probabilities than the irrelevant
predictors. Here we take $p=160$ predictors and 100 subsamples, fix
$\lambda_E$ at 0.1, and vary $\lambda_L$. The right panel of Figure
\ref{fig_stability} plots the average number of selected variables. One can
control the number of false positives by choosing the cutoff $\pi$.
Higher values of $\pi$ reduce both the number of false positives and
the number of true positives. Here an excellent balance is struck for
$\lambda_L$ between 0.1 and 0.2.

\subsection{Simulation examples 3 through 6}
\setcounter{example}{2}
To better assess the accuracy of the three new VDA methods, we now
present four three-class examples. In each example we generated 1000
predictors on 200 training observations and 1000 testing observations.
Unless stated to the contrary, all predictors were independent and
normally distributed with mean 0 and variance 1. Penalty tuning
constants were chosen by minimizing prediction error on the testing
data. We report average results from 50 random samples.
\begin{example}\label{exam3}
This is a multi-logit model with odds ratios
\[
\log\frac{\mathrm{Pr}(\mathrm{class}=l\vert x)}{\mathrm{Pr}(\mathrm{class}=3 \vert x)} =
\cases{
-x_{i1}-x_{i2}-x_{i3}+x_{i7}+x_{i8}, & \quad for class $1$, \cr
x_{i4}+x_{i5}+x_{i6}-x_{i7}-x_{i8}, & \quad for class $2$, \cr
1, & \quad for class $3$.}
\]
These ratios and the constraint $\sum_{i=1}^3 \mathrm{Pr}(\mathrm {class}=i)= 1$
determine class assignment. Obviously, only the first eight predictors
are relevant to classification.
\end{example}
\begin{example}\label{exam4}
In this example observations are equally distributed over classes. For
$j \le5$ the predictor $x_{ij}$ has mean $a_j$ with
\[
(a_1,a_2,a_3,a_4,a_5)  = \cases{
(0.5, 0.5, 1,0,0), & \quad for class $1$, \cr
(-0.5, -0.5,0,1,0), & \quad for class $2$, \cr
(0.5, -0.5,0,0,1), & \quad for class $3$.}
\]
The remaining predictors have mean 0 and are irrelevant to classification.
\end{example}

\begin{example}\label{exam5}
This example is the same as example 3 except that the first six
predictors are correlated with correlation coefficient $\rho=0.8$.
\end{example}

\begin{example}\label{exam6}
This example is the same as example 4 except that the first six
predictors are correlated with correlation coefficient $\rho=0.8$.
\end{example}

Table \ref{table_sim3} summarizes classification results for these
examples. In all instances VDA$_{\mathrm{LE}}$ and VDA$_{\mathrm E}$ show lower
prediction error rates than VDA$_{\mathrm L}$. In examples 3 and 4, where
predictors are independent, VDA$_{\mathrm{LE}}$ and VDA$_{\mathrm L}$ have much
higher false positive rates than VDA$_{\mathrm E}$. In defense of VDA$_{\mathrm{LE}}$,
 it has a lower prediction error and a higher true positive rate
than VDA$_{\mathrm E}$ in example 3. In examples 5 and 6, where predictors
are correlated, VDA$_{\mathrm{LE}}$ and VDA$_{\mathrm E}$ have much lower
prediction errors than VDA$_{\mathrm L}$; they also tend to better
VDA$_{\mathrm L}$ in variable selection.

\begin{table}
\tabcolsep=0pt
\caption{Comparison of VDA$_{\mathrm {LE}}$, VDA$_{\mathrm L}$, and VDA$_{\mathrm
E}$. Each line reports for 50 random replicates average prediction
error and 10\%, 50\%, and 90\% percentiles of the number of nonzero
variables and nonzero true variables selected. Standard errors for the
average prediction errors appear in parentheses}\label{table_sim3}
\begin{tabular*}{\textwidth}{@{\extracolsep{\fill}} l d{2.8} a{0.4} a{0.2} d{2.8} c a{0.2}@{} }
 \hline
\textbf{Method}
& \multicolumn{1}{c}{\textbf{Error (\%)}} &  \multicolumn{1}{c}{\textbf{\# Var}}  & \multicolumn{1}{c}{\textbf{\# True Var}}
& \multicolumn{1}{c}{\textbf{Error (\%)}} &  \multicolumn{1}{c}{\textbf{\# Var}} &  \multicolumn{1}{c@{}}{\textbf{\# True Var}}\\
\hline
&\multicolumn{3}{c}{Example 3} & \multicolumn{3}{c@{}}{Example 4}\\[-5pt]
& \multicolumn{3}{c}{\hrulefill} &\multicolumn{3}{c@{}}{\hrulefill}\\
VDA$_{\mathrm{LE}}$ & 36.13\ (0.36) & 17, 58, 219 & 7, 8, 8 & 31.65\ (0.31) &
5, 11, 64 & 5, 5, 5 \\
VDA$_{\mathrm L}$ & 37.57\ (0.34) & 20, 87, 264 & 7, 8, 8 & 34.05\ (0.31) &
8, 76, 214 & 5, 5, 5 \\
VDA$_{\mathrm E}$ & 37.27\ (0.38) & 13, 28, 65 & 6, 8, 8 & 32.11\ (0.33) & 5, 8, 21 &
5, 5, 5 \\[5 pt]
& \multicolumn{3}{c}{Example 5} & \multicolumn{3}{c@{}}{Example 6} \\[-5pt]
& \multicolumn{3}{c}{\hrulefill} &\multicolumn{3}{c@{}}{\hrulefill}\\
VDA$_{\mathrm{LE}}$ & 24.19\ (0.27) & 8, 14, 40 & 6, 7, 8 & 6.98\ (0.19) & 6, 11, 24 &
5, 5, 5 \\
VDA$_{\mathrm L}$ & 25.85\ (0.29) & 6, 30, 63 & 4, 6, 8 & 10.78\ (0.32) & 5, 12, 37 &
5, 5, 5 \\
VDA$_{\mathrm E}$ & 24.11\ (0.29) & 11, 19, 39 & 6, 7, 8 & 6.64\ (0.19) & 7, 19, 43 &
5, 5, 5 \\
\hline
\end{tabular*}
\end{table}

\section{Real data examples}\label{sec7}

\subsection{Overdetermined problems}

To test the performance of VDA models on real data, we first analyzed
four standard datasets (wine, glass, zoo, and lymphography) from the
UCI machine learning repository [\citet{murphy94}]. Table \ref{table_UCI}
compares the performance of the modified VDAs to the original VDA$_{\mathrm
R}$, linear discriminant analysis (LDA), quadratic discriminant
analysis (QDA), the $k$-nearest-neighbor method (KNN), one-versus-rest
binary support vector machines (OVR), classification and regression
trees (CART), random forest prediction, and multicategory support
vector machines (MSVM) [\citet{lee04}]. For all four datasets, the error
rates in the table are average misclassification rates based on 10-fold
cross-validation. We chose the penalty tuning constants for the various
VDA methods to minimize cross-validated errors over a one- or
two-dimensional grid. The entries of Table \ref{table_UCI} demonstrate
the effectiveness of VDA$_{\mathrm R}$ on small-scale problems. Our more
complicated method VDA$_{\mathrm{LE}}$ is a viable contender.

\begin{table}
\tabcolsep=0pt
\caption{Mean 10-fold cross-validated testing error rates for empirical
examples from the UCI data repository.
The triples beneath each dataset give in order the number of classes
$k$, the number of cases $n$, and the number of predictors $p$. NA
stands for not available} \label{table_UCI}
\begin{tabular*}{\textwidth}{@{\extracolsep{\fill}}l d{1.4} d{1.4} d{1.4} d{1.4}@{}}
\hline
& \multicolumn{1}{c}{\textbf{Wine}} & \multicolumn{1}{c}{\textbf{Glass}}&\multicolumn{1}{c}{ \textbf{Zoo}} & \multicolumn{1}{c@{}}{\textbf{Lymphography}}\\
\textbf{Method} &\multicolumn{1}{c}{\textbf{(3,\ 178,\ 13)}}& \multicolumn{1}{c}{\textbf{(6,\ 214,\ 10)}} &
\multicolumn{1}{c}{\textbf{(7,\ 101,\ 16)}}& \multicolumn{1}{c@{}}{\textbf{(4,\ 148,\ 18)}}\\
\hline
VDA$_{\mathrm R}$ & 0 & 0.2970 & 0.0182 & 0.0810 \\
VDA$_{\mathrm {LE}}$ & 0.0055 & 0.3267 & 0.0091 & 0.1210 \\
VDA$_{\mathrm L}$ & 0.0111 & 0.3357 & 0.0272 & 0.1277 \\
VDA$_{\mathrm E}$ & 0.0111 & 0.3420 & 0.0182 & 0.1620 \\[5 pt]
LDA & 0.0112 & 0.3972 & \multicolumn{1}{c}{NA} & 0.1486 \\
QDA & 0.0169 & \multicolumn{1}{c}{NA} & \multicolumn{1}{c}{NA} & \multicolumn{1}{c@{}}{NA} \\
KNN ($k=2$) & 0.2697 & 0.3084 & 0.0594 & 0.2432 \\
OVR & 0.0225 & 0.3178 & 0.0891 & 0.1486 \\
CART & 0.0899 & 0.4346 & 0.2475 & 0.2095 \\
Random forest & 0.0169 & 0.2009 & 0.0693 & 0.1621 \\
MSVM & 0.0169 & 0.3645 & \multicolumn{1}{c}{NA} & \multicolumn{1}{c@{}}{NA} \\
\hline
\end{tabular*}
\end{table}

\subsection{Underdetermined problems}

Our final examples are benchmark\break datasets for cancer diagnosis. These
public domain datasets are characterized by large numbers of predictors
and include the cancers: colon [\citet{alon99}], leukemia [\citet{golub99}],
 prostate [\citet{singh02}], brain [\citet{pomeroy00}], lymphoma
[\citet{alizadeh00}], and SRBCT [\citet{khan01}]. We compare our
classification results with those from three other studies
[\citet{li05}; \citet{statnikov05}; \citet{dettling04}]. Table \ref{table_cancer}
summarizes all findings. The cited error rates for BagBoost [\citet{dettling04}],
 Boosting [\citet{dettling03}], RanFor, SVM, nearest
shrunken centroids (PAM) [\citet{tibshirani02}], diagonal linear
discriminant analysis (DLDA) [\citet{tibshirani02}], and KNN appear in
[\citet{dettling04}]. The error rates in Table \ref{table_cancer} are
average misclassification rates based on 3-fold cross-validation. Again
we chose the penalty tuning constants for the various versions of VDA
by grid optimization. The error rates and training times listed in
Table \ref{table_cancer} are predicated on the selected tuning
constants.

\begin{table}
\tabcolsep=0pt
\caption{Threefold cross-validated testing errors (as percentages) for
six benchmark cancer datasets. The parenthesized triples for each
dataset give in order the number of categories $k$, the number of cases
$n$, and the number of predictors $p$. Column 2 and subsequent columns
report average testing error (standard error in parentheses), 10\%, 50\%,
 and 90\% percentiles of number of nonzero variables, and the average
training time in seconds over 50 random partitions. Execution times
apply to the entire dataset under the optimal tuning parameters
determined by cross-validation. All results for the non-VDA methods are
taken from the paper of Dettling (\protect\citeyear{dettling04})}\label{table_cancer}
\begin{tabular*}{\textwidth}{@{\extracolsep{\fill}} lccd{1.2}lccd{1.2}lccc@{}}
\hline
\textbf{Method}
& \multicolumn{1}{c}{\textbf{Error (\%)}} &  \multicolumn{1}{c}{\textbf{\# Var}} &  \multicolumn{1}{c}{\textbf{Time}}
 &&  \multicolumn{1}{c}{\textbf{Error (\%)}} &  \multicolumn{1}{c}{\textbf{\# Var}} &  \multicolumn{1}{c}{\textbf{Time}} && \multicolumn{1}{c}{\textbf{Error
 (\%)}}& \multicolumn{1}{c}{\textbf{\# Var}} &  \multicolumn{1}{c@{}}{\textbf{Time}} \\
\hline
& \multicolumn{3}{c}{Leukemia $(2,72,3571)$} && \multicolumn
{3}{c}{Colon $(2,62,2000)$ } && \multicolumn{3}{c@{}}{Prostate $(2,102,6033)$}
\\[-5pt]
&\multicolumn{3}{c}{\hrulefill} &&\multicolumn{3}{c@{}}{\hrulefill}&&\multicolumn{3}{c@{}}{\hrulefill}\\
VDA$_{\mathrm {LE}}$ & 1.56  & 18, 39, 74  & 0.50 & & 9.68  & 10, 27, 103& 0.15 & & 5.48& 16, 40, 53 & 1.15 \\
                     & (0.15)&             &      & &(0.55) &            &      & &(0.33)\\
VDA$_{\mathrm L}$    & 7.14  & 26, 30, 85  & 0.08 & & 14.26 & 19, 25, 147& 0.04 & & 9.83& 30, 36, 200 & 0.23 \\
                     &(0.62) &             &      & &(0.65) &            &      & &(0.56)\\
VDA$_{\mathrm E}$    & 3.02  & 42, 54, 179 & 0.45 & & 11.08 & 34, 42, 213& 0.12 & & 6.76& 47, 57, 366 & 0.85 \\
                     & (0.28)&             &      & &(0.52) &            &      & &(0.41)\\[3pt]
BagBoost             & 4.08  &             &      & & 16.10 &            &      & &7.53 \\
Boosting             & 5.67  &             &      & & 19.14 &            &      & & 8.71 \\
RanFor               & 1.92  &             &      & & 14.86 &            &      & &9.00 \\
SVM                  & 1.83  &             &      & & 15.05 &            &      & &7.88 \\
PAM                  & 3.75  &             &      & & 11.90 &            &      & &16.53 \\
DLDA                 & 2.92  &             &      & & 12.86 &            &      & &14.18 \\
KNN                  & 3.83  &             &      & & 16.38 &            &      & &10.59 \\ [5pt]
& \multicolumn{3}{c}{Lymphoma $(3,62,4026)$} && \multicolumn
{3}{c}{SRBCT $(4,63,2308)$} && \multicolumn{3}{c@{}}{Brain $(5,42,5597)$}\\[-5 pt]
&\multicolumn{3}{c}{\hrulefill}& &\multicolumn{3}{c@{}}{\hrulefill}&&\multicolumn{3}{c@{}}{\hrulefill}\\
VDA$_{\mathrm {LE}}$ & 1.66  &   39, 69, 97 & 1.47& & 1.58& 45, 60, 94 &
1.78 && 23.80 & 52, 78, 98 & 4.39 \\
& (0.27)& & &&(0.77)& & & &(1.54)\\
VDA$_{\mathrm L}$ & 14.36 & 39, 53, 86 & 0.12& & 9.52 & 43, 53, 65 &
0.11 && 48.86& 46, 57, 66 & 0.38 \\
& (0.97)& & &&(1.14)& & & &(1.43)\\
VDA$_{\mathrm E}$ & 3.25 & 80, 91, 128 & 2.01& & 1.58 & 58, 70, 106
& 1.70& & 30.44 & 70, 85, 100 & 6.43 \\
& (0.38)& & &&(0.92)& & && (1.76)\\[3pt]
BagBoost & 1.62&& & &1.24&& && 23.86 \\
Boosting & 6.29&& & &6.19&& && 27.57 \\
RanFor & 1.24&& & &3.71&& & &33.71 \\
SVM & 1.62&& & &2.00&& & &28.29 \\
PAM & 5.33&& & &2.10&& & &25.29 \\
DLDA & 2.19&& && 2.19&& & &28.57 \\
KNN & 1.52&& & &1.43&& & &29.71 \\
\hline
\end{tabular*}
\end{table}

\begin{figure}

\includegraphics{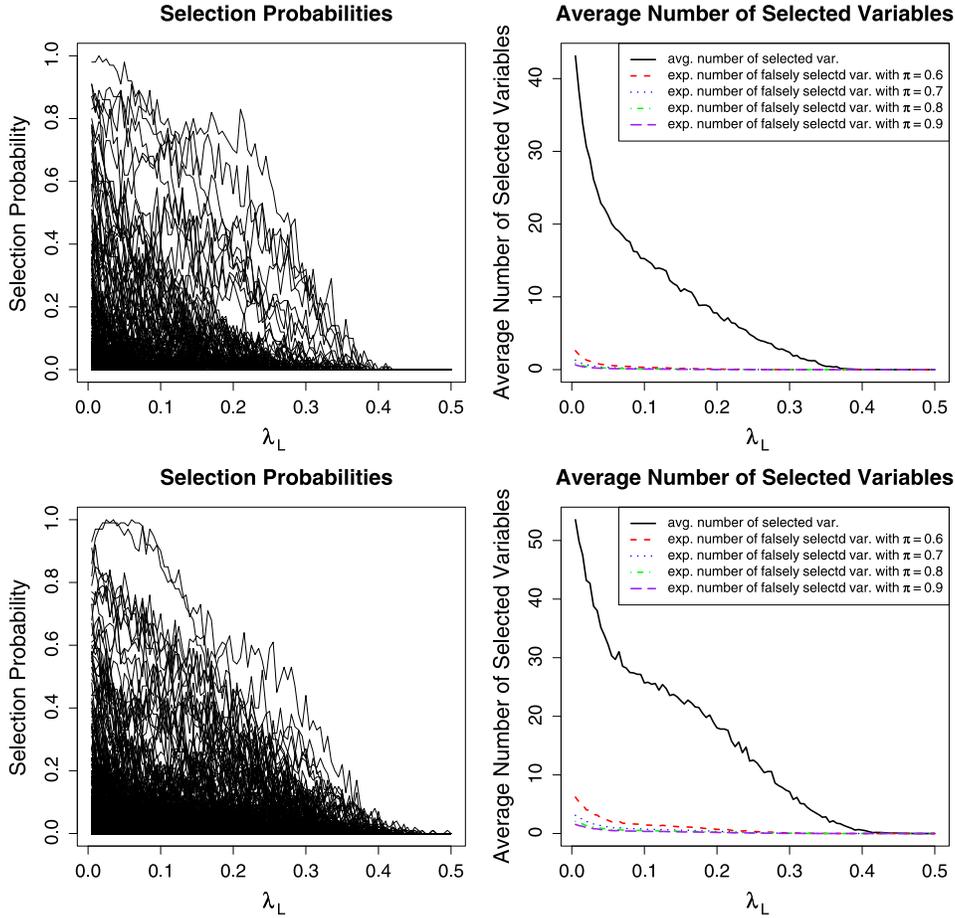}

\caption{Stability selection with VDA$_{\mathrm{LE}}$ for leukemia (upper
panel with $k=2$) and SRBCT (lower panel with $k=3$) data. The left
panels plot the empirical selection probabilities of all predictors
over 100 subsamples as a function of $\lambda_L$ for $\lambda_E$ fixed
at 0.1. The right panel plots the average number of selected predictors
(black solid line) and the expected number of falsely selected
predictors for different values of the cutoff $\pi$.}\label{fig_stability_data}
\end{figure}

Inspection of Table \ref{table_cancer} suggests that VDA$_{\mathrm {LE}}$ may
be superior to the popular classifiers listed. Although very fast,
VDA$_{\mathrm L}$ is not competitive with VDA$_{\mathrm {LE}}$; VDA$_{\mathrm E}$
performs well but falters on the lymphoma and brain examples. Owing to
the large number of predictors, application of VDA$_{\mathrm R}$ is
impractical in these examples. We also applied stability selection to
the leukemia and SRBCT data. As Figure~\ref{fig_stability_data}
demonstrates, the expected number of false positives is small across a
range of cutoff values $\pi$.

\section{Discussion}\label{sec8}

As one of the most important branches of applied statistics,
discriminant analysis continues to attract the attention of
theoreticians. Although the flux of new statistical demands and ideas
has not produced a clear winner among the competing methods, we hope to
have convinced readers that VDA and its various modifications are
competitive. It is easy to summarize the virtues of VDA in four words:
parsimony, robustness, speed, and symmetry. VDA$_{\mathrm R}$ excels in
robustness and symmetry but falls behind in parsimony and speed. We
recommend it highly for problems with a handful of predictors.
VDA$_{\mathrm E}$ excels in robustness, speed, and symmetry. On
high-dimensional problems it does not perform
quite as well as VDA$_{\mathrm {LE}}$, which sacrifices a little symmetry for
extra parsimony. Apparently, VDA$_{\mathrm L}$ puts too high a premium on
parsimony at the expense of symmetry.

Our Euclidean penalties tie together the parameters corresponding to a
single predictor. Some applications may require novel ways of grouping
predictors. For example in cancer diagnosis, genes in the same
biological pathway could be grouped. If reliable grouping information
is available, then one should contemplate adding further Euclidean
penalties [\citet{wu08}]. If other kinds of structures exist, one should
opt for different penalty functions. For example, \citet{yuan09} and
\citet{wuzou08} discuss the problem of how to retain hierarchical
structures in variable selection using the nonnegative garrote [\citet{breiman95}].

The class vertices in VDA are symmetrically distributed over the
surface of the unit ball. When categories are ordered or partially
ordered, equidistant vertices may not be optimal. The question of how
to incorporate order constraints deserves further investigation. The
simplest device for three ordered categories is to identify them with
the points $-1$, 0, and 1 on the line.

Future applications of discriminant analysis will confront even
larger\break
datasets. Computing times are apt to balloon out of control unless the
right methods are employed. Cyclic coordinate descent has proved to be
extraordinarily fast when coupled with lasso or Euclidean penalties.
The same speed advantages are seen in lasso penalized regression and
generalized linear models. Further gains in speed may well come in
parallel computing. Statisticians have been slow to plunge into
parallel computing because of the extra programming effort required and
the lack of portability across computing platforms. It is not clear how
best to exploit parallel computing with VDA.

Stability selection as sketched by \citet{meinshausen09} appears to work
well with VDA. In our simulated example, it eliminates virtually all
irrelevant predictors while retaining the true predictors. For the
cancer data, the true predictors are unknown; it is encouraging that
the expected number of false positives is very low. Because stability
selection requires repeated subsampling of the data, users will pay a
computational price. This cost is not excessive for VDA, and we highly
recommend stability selection. In our view it will almost certainly
replace cross-validation in model selection.

The theoretical underpinnings of VDA and many other methods of
discriminant analysis are weak. We prove Fisher consistency here, but
more needs to be done. For instance, it would be reassuring if someone
could vindicate our intuition that shrinkage is largely irrelevant to
classification by VDA. Although it is probably inevitable that
statistical practice will outrun statistical theory in discriminant
analysis, ultimately there is no stronger tether to reality than a good
theory. Of course, a bad or irrelevant theory is a waste of time.

\begin{supplement}[id=suppA]
  \sname{Supplementary File}
  \stitle{Proof of Proposition 1}
  \slink[doi]{10.1214/10-AOAS345SUPP} 
  \sdatatype{.pdf}
  \slink[url]{http://lib.stat.cmu.edu/aoas/345/supplement.pdf}
  \sdescription{We prove Fisher consistency of $\varepsilon$-insensitive loss in this paper.}
\end{supplement}

\printaddresses

\end{document}